\documentclass[11pt]{article}
\usepackage{amssymb}
\usepackage{amsmath}
\usepackage{amsfonts}
\usepackage{latexsym}
\usepackage{graphicx}

\newtheorem{theorem}{Theorem}
\newtheorem{definition}[theorem]{Definition}
\newtheorem{lemma}[theorem]{Lemma}
\newtheorem{corollary}[theorem]{Corollary}
\newtheorem{proposition}[theorem]{Proposition}
\newtheorem{conjecture}{Conjecture}

\newcommand{\qed}{$\square$}

\newenvironment{proof}{%
  \noindent{\it Proof.\ }}{%
  \hspace*{\fill}\qed
  \vspace{2ex}}

\newcommand{\Hamming}[1]{\mathrm{Ham}(#1)}
\newcommand{\tr}{\mathrm{tr}}
\newcommand{\neighbor}{\mathrm{neighbor}}
\newcommand{\prob}{\mathop{\it Prob}}
\newcommand{\nominal}{\mathop{\it Time}\nolimits_{\rm nom}}
\newcommand{\real}{\mathop{\it Time}\nolimits_{\rm real}}
\newcommand{\ab}{\mathrm{absorb}}
\newcommand{\bra}[1]{\langle #1 |}
\newcommand{\ket}[1]{| #1 \rangle}

\title{\Large\bf Analysis of Absorbing Times of Quantum Walks}
\author{
  {\large $\mbox{\bf Tomohiro Yamasaki}^{\ast}$}\\
  {\tt \hspace*{-2ex} yamasaki@is.s.u-tokyo.ac.jp \hspace*{-2ex}} 
  \and
  {\large $\mbox{\bf Hirotada Kobayashi}^{\ast\dagger}$}\\
  {\tt \hspace*{-2ex} hirotada@qci.jst.go.jp \hspace*{-2ex}}
  \and
  {\large $\mbox{\bf Hiroshi Imai}^{\dagger\ddagger}$}\\
  {\tt \hspace*{-2ex} imai@is.s.u-tokyo.ac.jp \hspace*{-2ex}} 
}

\date{}

\begin{document}
\maketitle
\thispagestyle{plain}
\pagestyle{plain}

\begin{center}
{\large       ${}^{\ast}$Department of Information Science\\
        Graduate School of Science\\
        The University of Tokyo\\
        7-3-1 Hongo, Bunkyo-ku, Tokyo 113-0033, Japan}

\vspace{5mm}
{\large        ${}^{\dagger}$Quantum Computation and Information Project\\
        Exploratory Research for Advanced Technology\\
        Japan Science and Technology Corporation\\
        5-28-3 Hongo, Bunkyo-ku, Tokyo 113-0033, Japan}

\vspace{5mm}
{\large       ${}^{\ddagger}$Department of Computer Science\\
        Graduate School of Information Science and Technology\\
        The University of Tokyo\\
        7-3-1 Hongo, Bunkyo-ku, Tokyo 113-0033, Japan}

\vspace{8mm}
{\large 28 April 2003}
\end{center}
\vspace{3mm}


\begin{abstract}
Quantum walks are expected to provide useful algorithmic tools
for quantum computation. 
This paper introduces
absorbing probability and time of quantum walks
and gives both numerical simulation results and theoretical analyses
on Hadamard walks on the line
and symmetric walks on the hypercube
from the viewpoint of absorbing probability and time.
\end{abstract}

\maketitle

\section{Introduction}
Random walks, or often called Markov chains, on graphs have found a number of 
applications in various fields,
not only in natural science such as physical systems
and mathematical modeling of life phenomena
but also in social science such as financial systems.
Also in computer science,
random walks have been applied to various problems such as 2-SAT,
approximation of the permanent~\cite{JerSin89SIComp,JerSinVig01STOC},
and estimation of the volume of convex bodies~\cite{DyeFriKan91JACM}.
Sch\"{o}ning's elegant algorithm for 3-SAT~\cite{Sch99FOCS}
and its improvement~\cite{HofSchSchWat02STACS}
are also based on classical random walks.
Moreover, one of the most advantageous points of classical random walks
as a useful algorithmic tool is
that they use only simple local transitions
to obtain global properties of the instance.

Thus, it is natural to consider
quantum generalization of classical random walks,
which may be very useful in constructing efficient quantum algorithms,
for which only a few general algorithmic tools have been developed,
including Fourier sampling and amplitude amplification.
There have been considered two types of quantum walks:
one is a discrete-time walk discussed by
Watrous~\cite{Wat01JCSS},
Ambainis, Bach, Nayak, Vishwanath, and Watrous~\cite{AmbBacNayVisWat01STOC},
Aharonov, Ambainis, Kempe, and Vazirani~\cite{AhaAmbKemVaz01STOC},
and Moore and Russell~\cite{MooRus02RANDOM},
and the other is a continuous-time one by
Farhi and Gutmann~\cite{FarGut98PRA},
Childs, Farhi, and Gutmann~\cite{ChiFarGut02JQIP},
and Moore and Russell~\cite{MooRus02RANDOM}.
All of these studies demonstrate that the behaviors of quantum walks are quite different from classical ones.
In the theory of classical random walks,
one of the important measures is the {\em mixing time\/},
which is the time necessary to have the probability distribution of the particle
be sufficiently close to the stationary distribution.
Unfortunately, Aharonov, Ambainis, Kempe, and Vazirani~\cite{AhaAmbKemVaz01STOC}
showed a rather negative result that
discrete-time quantum walks can be at most polynomially faster
than classical ones in view of mixing time.
As for continuous-time quantum walks,
Farhi and Gutmann~\cite{FarGut98PRA} and
Childs, Farhi, and Gutmann~\cite{ChiFarGut02JQIP} claimed that
quantum walks propagate exponentially faster than classical ones.
Furthermore, a recent result
by Child, Cleve, Farhi, Deotto, and Spielman~\cite{ChiCleDeoFarGutSpi03STOC}
has shown an example of a black-box problem in which
continuous time quantum walks can provide an exponential speedup
to all classical algorithms (not necessarily limited to random walks).

This paper focuses on the discrete-time type of quantum walks
and introduces two new criteria for propagation speed of discrete-time quantum walks: {\em absorbing probability\/} and {\em absorbing time\/}.
A number of properties of quantum walks are investigated in view of these criteria.
In particular, the behaviors of Hadamard walks on the line
and symmetric walks on the hypercube
are discussed through the results of numerical simulation experiments.
In our simulations, quantum walks on the hypercube appear exponentially faster
than classical ones in absorbing time under certain situations.
More precisely, such a speedup would happen in the case
the absorbing vertex is located ``near'' to the antipodal vertex of
the initial vertex.
Several theoretical analyses are also given on classical and quantum symmetric walks
on the hypercube.

The remainder of this paper is organized as follows.
Section~\ref{Section: Definitions} reviews
the formal definition of (discrete-time) quantum walks,
and introduces new criteria for propagation speed of quantum walks.
Section~\ref{Section: Hadamard Walks}
and Section~\ref{Section: Symmetric Walks}
deal with a number of numerical simulations on Hadamard walks on the line
and symmetric walks on the hypercube, respectively.
Section~\ref{Section: Theoretical Analyses} gives
several theoretical analyses on symmetric walks on the hypercube.
Finally, we conclude with Section~\ref{Section: Conclusions},
which summarizes our results.

\section{Definitions}
\label{Section: Definitions}

In this section, we give a formal definition of quantum walks.
As mentioned in the previous section,
there exist two types of quantum generalizations of classical random walks:
discrete-time quantum walks and continuous-time quantum walks. 
Here we only give a definition of discrete-time ones,
which this paper treats.
We also introduce new criteria of absorbing probability and absorbing time
for propagation speed of quantum walks.

\subsection{Discrete-Time Walks}

In direct analogy to classical random walks, one may try to define quantum
walks as follows: at each step, the particle moves in all directions
with equal amplitudes. However, such a walk is impossible
for a discrete-time model
since the evolution of whole quantum system would not be always unitary.
Indeed Meyer~\cite{Mey96JStatPhy} proved the impossibility
of such discrete-time quantum walks on a lattice.
Here we review the definition of discrete-time quantum walks
on graphs according to~\cite{AhaAmbKemVaz01STOC}.

\begin{figure}[t]
\setlength{\unitlength}{1mm}
\begin{center}
\begin{picture}(100,40.75)
  \put(10, 0){\circle*{1.5}}
  \put(10,30){\circle*{1.5}}
  \put(40,30){\circle*{1.5}}
  \put(40, 0){\circle*{1.5}}
  \put(17.5, 7.5){\circle*{1.5}}
  \put(17.5,22.5){\circle*{1.5}}
  \put(32.5,22.5){\circle*{1.5}}
  \put(32.5, 7.5){\circle*{1.5}}
  \put(10, 0){\vector(0,1){30}}
  \put(10,30){\vector(1,0){30}}
  \put(40,30){\vector(0,-1){30}}
  \put(40, 0){\vector(-1,0){30}}
  \put(17.5, 7.5){\vector(0,1){15}}
  \put(17.5,22.5){\vector(1,0){15}}
  \put(32.5,22.5){\vector(0,-1){15}}
  \put(32.5, 7.5){\vector(-1,0){15}}
  \put(10, 0){\vector(1,1){7.5}}
  \put(10,30){\vector(1,-1){7.5}}
  \put(40,30){\vector(-1,-1){7.5}}
  \put(40, 0){\vector(-1,1){7.5}}
  \put(17.5,7.5){\vector(-1,-1){7.5}}
  \put(17.5,22.5){\vector(-1,1){7.5}}
  \put(32.5,22.5){\vector(1,1){7.5}}
  \put(32.5,7.5){\vector(1,-1){7.5}}
  \put(7,-3){\makebox(0,0){$g^3b$}}
  \put(7,33){\makebox(0,0){$h$}}
  \put(43,33){\makebox(0,0){$gh$}}
  \put(43,-3){\makebox(0,0){$g^2h$}}
  \put(20.5,10.5){\makebox(0,0){$g^3$}}
  \put(20.5,19.5){\makebox(0,0){$e$}}
  \put(29.5,19.5){\makebox(0,0){$g$}}
  \put(29.5,10.5){\makebox(0,0){$g^2$}}
  \put(25,-1.5){\makebox(0,0){\scriptsize $g$}}
  \put(25,31.5){\makebox(0,0){\scriptsize $g$}}
  \put(8.5,15){\makebox(0,0){\scriptsize $g$}}
  \put(41.5,15){\makebox(0,0){\scriptsize $g$}}
  \put(25,6){\makebox(0,0){\scriptsize $g$}}
  \put(25,24){\makebox(0,0){\scriptsize $g$}}
  \put(16,15){\makebox(0,0){\scriptsize $g$}}
  \put(34,15){\makebox(0,0){\scriptsize $g$}}
  \put(13.5,5.5){\makebox(0,0){\scriptsize $h$}}
  \put(36.5,5.5){\makebox(0,0){\scriptsize $h$}}
  \put(13.5,24.5){\makebox(0,0){\scriptsize $h$}}
  \put(36.5,24.5){\makebox(0,0){\scriptsize $h$}}
  \put(60, 0){\circle*{1.5}}
  \put(60,30){\circle*{1.5}}
  \put(90,30){\circle*{1.5}}
  \put(90, 0){\circle*{1.5}}
  \put(67.5, 7.5){\circle*{1.5}}
  \put(67.5,22.5){\circle*{1.5}}
  \put(82.5,22.5){\circle*{1.5}}
  \put(82.5, 7.5){\circle*{1.5}}
  \put(60,0){\framebox(30,30){}}
  \put(67.5,7.5){\framebox(15,15){}}
  \put(60, 0){\line(1,1){7.5}}
  \put(60,30){\line(1,-1){7.5}}
  \put(90,30){\line(-1,-1){7.5}}
  \put(90, 0){\line(-1,1){7.5}}
  \put(58, 3){\makebox(0,0){\small 1}}
  \put(63,32){\makebox(0,0){\small 1}}
  \put(92,27){\makebox(0,0){\small 1}}
  \put(87,-2){\makebox(0,0){\small 1}}
  \put(62.5, 2.5){\makebox(0,0){\small 2}}
  \put(62.5,27.5){\makebox(0,0){\small 2}}
  \put(87.5,27.5){\makebox(0,0){\small 2}}
  \put(87.5, 2.5){\makebox(0,0){\small 2}}
  \put(63,-2){\makebox(0,0){\small 3}}
  \put(58,27){\makebox(0,0){\small 3}}
  \put(87,32){\makebox(0,0){\small 3}}
  \put(92, 3){\makebox(0,0){\small 3}}
  \put(65.5, 9.5){\makebox(0,0){\small 1}}
  \put(69.5,24.5){\makebox(0,0){\small 1}}
  \put(84.5,20.5){\makebox(0,0){\small 1}}
  \put(80.5, 5.5){\makebox(0,0){\small 1}}
  \put(65.5, 5.5){\makebox(0,0){\small 2}}
  \put(65.5,24.5){\makebox(0,0){\small 2}}
  \put(84.5,24.5){\makebox(0,0){\small 2}}
  \put(84.5, 5.5){\makebox(0,0){\small 2}}
  \put(69.5, 5.5){\makebox(0,0){\small 3}}
  \put(65.5,20.5){\makebox(0,0){\small 3}}
  \put(80.5,24.5){\makebox(0,0){\small 3}}
  \put(84.5, 9.5){\makebox(0,0){\small 3}}
\end{picture}
\end{center}
\vspace*{-1mm}
\caption{An example of a Cayley graph and its labeling
($g^4 = e, h^2 = e, g \circ h = h \circ g$).}
\label{Figure: Cayley graph}
\end{figure}
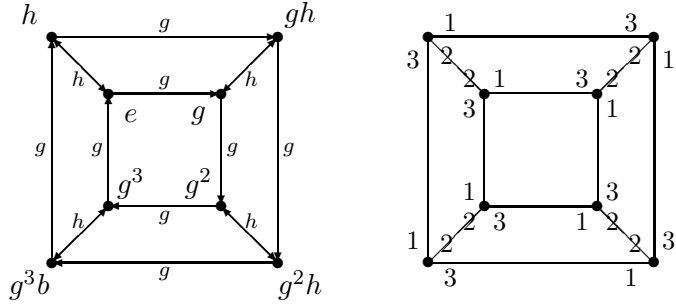

Let $G = (V,E)$ be a graph where $V$ is a set of vertices and $E$ is a set
of edges. For theoretical convenience, we assume that $G$ is
$d$-regular (i.e. each vertex of the graph $G$ is connected to exactly
$d$ edges).
For each vertex $v \in V$, label each edge
connected to this vertex with a number between $1$ and $d$
such that, for each $a \in \{1, \ldots, d\}$,
the directed edges labeled $a$ form a permutation.
In other words, for each vertex $v \in V$, not only label
each outgoing edge $(v,w) \in E$ with a number between $1$ and $d$ but
also label each incoming edge $(w,v) \in E$ with a number between $1$
and $d$, where $w$ is a vertex in $V$ and $w \neq v$. In the case that
$G$ is a Cayley graph, the labeling of a directed edge simply
corresponds to the generator associated with the edge.
An example of a Cayley graph and its labeling are
in Figure~\ref{Figure: Cayley graph}.

Let $\mathcal{H}_V$ be the Hilbert space spanned by $\{ \ket{v} \mid v \in V \}$, and let $\mathcal{H}_A$ be the auxiliary Hilbert space
spanned by $\{ \ket{a} \mid 1 \le a \le d \}$.
We think of this auxiliary Hilbert space as a ``coin space''.

\begin{definition}[\cite{AhaAmbKemVaz01STOC}]
Let $C$ be a unitary operator on $\mathcal{H}_A$ which we think of as
a ``coin-tossing operator'', and let a shift operator $S$ on
$\mathcal{H}_A \otimes \mathcal{H}_V$ be defined as
$S\ket{a,v} = \ket{a,w}$ where $w$ is the $a$th neighbor of vertex $v$.
One step of transition of the quantum walk on the graph $G$ is defined
by a unitary operator $W = S(C \otimes I_V)$ where $I_V$ is the
identity operator on $\mathcal{H}_V$. We call such a walk a discrete-time
quantum walk.
\end{definition}

More generally, we relax the restriction on the exact form of the quantum
walk operator $W$ and define general quantum walks such that $W$
respects only the structure of the graph $G$.
In other words, we require that,
in the superposition $W \ket{a,v}$
for each $a \in \{1, \ldots, d\}$ and $v \in V$,
only basis states $\ket{a^\prime,v^\prime}$ for
$a^\prime \in \{1, \ldots, d\}, v^\prime \in \neighbor(v) \cup \{v\}$
have non-zero amplitudes,
where $\neighbor(v)$ is the set of vertices adjacent to $v$.
Thus the graph $G$ does not need to be $d$-regular
and the particle at $v$ in the quantum walk
moves to one of the vertices adjacent to $v$ or stays at $v$ in one step.

\subsection{Criteria of Propagation Speed}

One of the properties of random or quantum walks we investigate is
how fast they spread over a graph.
In order to evaluate it, a criterion
``{\em mixing time\/}'' has been considered traditionally.
However, Aharonov, Ambainis, Kempe, and Vazirani~\cite{AhaAmbKemVaz01STOC}
showed that, in view of mixing time,
discrete-time quantum walks can be at most polynomially faster
than classical ones.
To seek the possibility of quantum walks being advantageous to classical ones,
this paper introduces another two new criteria of
propagation speed of quantum walks: {\em absorbing probability\/} and {\em absorbing time\/}.
Let us consider the set of {\em absorbing vertices\/}
such that the particle is absorbed if it reaches a vertex in this set.
This is done by
considering a measurement
that is described by
the projection operators over
$\mathcal{H}_A \otimes \mathcal{H}_V$:
$P = I_A \otimes \sum_{v \in \ab(V)}\ket{v}\bra{v}$ and
$P^\prime = I_A \otimes \sum_{v \not\in \ab(V)}\ket{v}\bra{v}$,
where $\ab(V) \subset V$ is the set of the absorbing vertices and $I_A$
is the identity operator over $\mathcal{H}_A$.

By using density operators,
one step of transition of a
discrete-time quantum walk can be described by a completely positive (CP) map $\Lambda$ as
follows:
\[
\Lambda: \rho \mapsto \rho^\prime = IP\rho PI + WP^\prime\rho P^\prime W^{\dagger},
\]
where $\rho$ and $\rho^\prime$ are density operators over the entire system
of the quantum walk,
$W$ is a unitary operator corresponding to one step of transition of the quantum walk,
and $I$ is the identity operator on
$\mathcal{H}_A \otimes \mathcal{H}_V$.

Now, we give a formal definition of absorbing probability.
\begin{definition}
Absorbing probability of a quantum walk is the probability that the 
particle which starts at the initial vertex
eventually reaches one of the absorbing vertices.
More formally, it is defined as
\[
\prob = \sum_{t=0}^\infty p(t),
\]
where $p(t)$ is the probability that
the particle reaches one of the absorbing vertices at time $t$ for the first time.
\end{definition}

Let $\ket{\psi_0}$ be the initial state of the system, and let
$\rho_0 = \ket{\psi_0}\bra{\psi_0}$.
Then $p(t) = \tr(\Lambda^t(\rho_0)P) - \tr(\Lambda^{t-1}(\rho_0)P)$
since
$\tr(\Lambda^t(\rho_0)P)$ and $\tr(\Lambda^{t-1}(\rho_0)P)$ are the
probabilities that the particle is absorbed by the time $t$ and $t-1$,
respectively.
Here we allowed a little abuse of a notation $\Lambda$
such that $\Lambda^2 (\rho) = \Lambda (\Lambda (\rho))$,
${\Lambda^3 (\rho) = \Lambda (\Lambda^2 (\rho))}$,
\ldots, $\Lambda^t (\rho) = \Lambda (\Lambda^{t-1} (\rho))$, and so on.

Next, we give a formal definition of {\em nominal absorbing time\/}.
\begin{definition}
Nominal absorbing time of a quantum walk is the expected time
to have the particle which starts at the initial vertex
eventually reach one of the absorbing vertices.
More formally, it is defined as
\[
  \nominal = \sum_{t=0}^\infty t p(t),
\]
where $p(t)$ is the probability that
the particle reaches one of the absorbing vertices at time $t$ for the first time.
\end{definition}

Even though nominal absorbing time is bounded polynomial in the size of the graph instance,
the quantum walk may not be efficient from a viewpoint of
computational complexity if absorbing probability is exponentially small,
because we need repeating the walk $O(1/p)$ times
to have a total absorption probability of $O(1)$
in the case the probability of absorption is $p$.
Thus we cannot regard nominal absorbing time as a criterion of how fast a
quantum walk spreads over a graph.
Instead, we use the following {\em real absorbing time\/}
as a criterion of the propagation speed of quantum walks.
\begin{definition}
Real absorbing time of a quantum walk is defined as
\[
  \real = \frac{\nominal}{\prob}
   = \frac{\sum_{t=0}^\infty t p(t)}{\sum_{t=0}^\infty p(t)},
\]
 where $p(t)$ is the probability that
 the particle reaches one of the absorbing vertices at time $t$ for the first time.
\end{definition}

In what follows,
we may refer to this real absorbing time as absorbing
time in short, when it is not confusing.

\section{Hadamard Walks on the Line}
\label{Section: Hadamard Walks}

\subsection{The Model}

We start with the model of discrete-time quantum walks on the line,
which was introduced and discussed by Ambainis, Bach, Nayak, Vishwanath, and Watrous~\cite{AmbBacNayVisWat01STOC}.
Note that the line can be regarded as a Cayley graph of the Abelian
group $\mathbb{Z}$ with generators $+1$ and $-1$,
denoted by ``$\mathrm{R}$'' and ``$\mathrm{L}$'', respectively.
Thus, the shift operator $S$ is defined as
\[
\begin{array}{lcl}
  S \ket{\mathrm{R},v} &=& \ket{\mathrm{R},v+1},\\
  S \ket{\mathrm{L},v} &=& \ket{\mathrm{L},v-1},
\end{array}
\]
where $v \in \mathbb{Z}$ is the vertex at which the particle is located.

Recall that the particle doing a classical random walk determines a
direction to move randomly at each step.
In the quantum case, therefore,
it is quite natural to set a coin-tossing operator $C$ as the Hadamard operator
\[
H = \frac{1}{\sqrt{2}}
      \left(
        \begin{array}{rr}
          1 & 1\\
          1 & -1
        \end{array}
      \right).
\]
We call this quantum walk an {\em Hadamard walk on the line\/}.
A number of properties were shown in~\cite{AmbBacNayVisWat01STOC}
on this model,
including several different aspects of Hadamard walks from their classical counterparts.

This paper considers the following process for some fixed $m$~\cite{Yam02Master},
which is slightly different from the model above.

\begin{enumerate}
\item
  Initialize the system in the state $\ket{\mbox{R}, 0}$,
  that is, let the particle be located at vertex $0$ at time $0$,
  with the chirality being $\mbox{R}$.
\item
  \begin{enumerate}
  \item
    Apply $W = S(H \otimes I_\mathbb{Z})$.
  \item
    Measure the system according to $\{\Pi_m, \Pi_m^\prime\}$
    such that $\Pi_m = I_2 \otimes \ket{m}\bra{m}$ and
    $\Pi_m^\prime = I_2 \otimes (I_\mathbb{Z} - \ket{m}\bra{m})$,
    where $I_2$ is the two-dimensional identity operator
    (i.e. measure the system to observe whether the particle
    is located at the vertex $m$ or not).
  \end{enumerate}
\item
  If the particle is observed to be located at the vertex $m$ after the 
  measurement, the process terminates, otherwise it repeats Step 2.
\end{enumerate}

\subsection{Numerical Simulations}

\begin{figure}[t]
\begin{center}
\includegraphics[height=5.75cm]{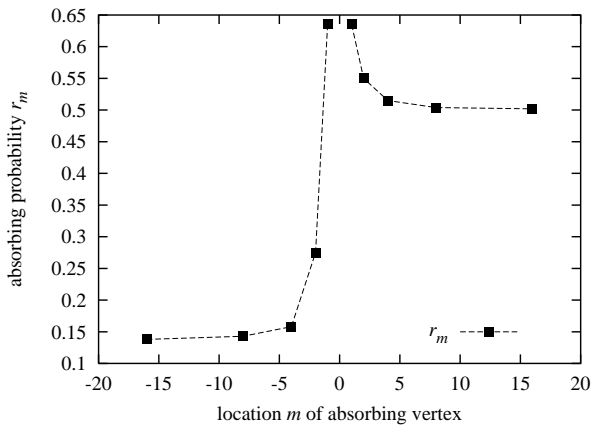}
\end{center}
\vspace*{-5mm}
\caption{Absorbing probability $r_m$ for Hadamard walks.}
\label{Figure: r_m}
\end{figure}

Let $r_m$ be absorbing probability that the particle is eventually
absorbed by the boundary at the vertex $m$.
Figure~\ref{Figure: r_m} illustrates the relation between $m$ (in $x$-axis)
and $r_m$ (in $y$-axis).
From Figure~\ref{Figure: r_m},
we conjecture that $\lim_{m\to\infty}r_m = 1/2$.

\begin{figure}[t]
\begin{center}
\includegraphics[height=5.75cm]{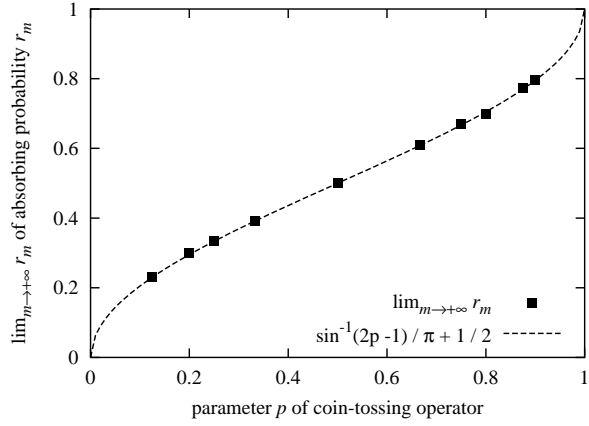}
\end{center}
\vspace*{-5mm}
\caption{Relation between $p$ and $\lim_{m \to +\infty}r_m$ for
generalized Hadamard walks.}
\label{Figure: r_m2}
\end{figure}

Next, let us consider a more general model of Hadamard walks
whose coin-tossing operator is defined as the matrix
\[
H_p = \left(
        \begin{array}{cc}
          \sqrt{\mathstrut p} & \sqrt{\mathstrut 1-p}\\
          \sqrt{\mathstrut 1-p} & -\sqrt{\mathstrut p}
        \end{array}
      \right),
\]
instead of $H$.
Figure~\ref{Figure: r_m2} illustrates the relation between $p$ (in $x$-axis)
and $\lim_{n\to+\infty}r_m$ (in $y$-axis)
for these generalized Hadamard walks.
Although we have not yet derived a closed form for $r_m$,
we conjecture the following.
\begin{conjecture}
For a generalized Hadamard walk with a coin-tossing operator $H_p$,
\[
\displaystyle\lim_{m\to\infty}r_m =
\frac{\sin^{-1}(2p-1)}{\pi}+\frac{1}{2}.
\]
\end{conjecture}

\noindent
{\it Remark}~~
This conjecture was proved very recently by Bach, Coppersmith,
Goldschen, Joynt, and Watrous~\cite{BacCopGolJoyWat02quant-ph}.

\section{Symmetric Walks on the $n$-Dimensional Hypercube}
\label{Section: Symmetric Walks}

\subsection{The Model}

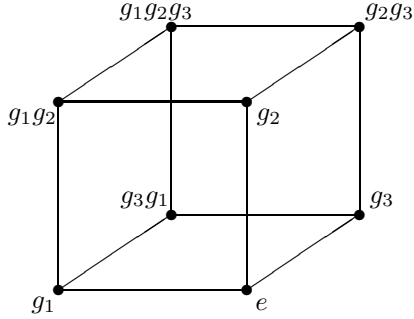
\begin{figure}[t]
  \setlength{\unitlength}{0.5mm}
  \begin{center}
    \begin{picture}(100,75)(-5,5)
      \put(0,0){\circle*{3}}
      \put(50,0){\circle*{3}}
      \put(0,50){\circle*{3}}
      \put(50,50){\circle*{3}}
      \put(30,20){\circle*{3}}
      \put(80,20){\circle*{3}}
      \put(30,70){\circle*{3}}
      \put(80,70){\circle*{3}}
      \put(0,0){\framebox(50,50){}}
      \put(30,20){\framebox(50,50){}}
      \put(0,0){\line(3,2){30}}
      \put(50,0){\line(3,2){30}}
      \put(0,50){\line(3,2){30}}
      \put(50,50){\line(3,2){30}}
      \put(-4,-4){\makebox(0,0){\small $g_1$}}
      \put(54,-4){\makebox(0,0){\small $e$}}
      \put(56,46){\makebox(0,0){\small $g_2$}}
      \put(86,24){\makebox(0,0){\small $g_3$}}
      \put(-7,46){\makebox(0,0){\small $g_1g_2$}}
      \put(88,74){\makebox(0,0){\small $g_2g_3$}}
      \put(23,24){\makebox(0,0){\small $g_3g_1$}}
      \put(26,74){\makebox(0,0){\small $g_1g_2g_3$}}
    \end{picture}
  \end{center}
  \caption{The $3$-dimensional hypercube.
    Each vertex can be regarded as a binary representation of length $3$
    (e.g.$g_1 \simeq (100)_2$ or $g_2g_3 \simeq (011)_2$).}
\label{Figure: hypercube}
\end{figure}

Next we consider the discrete-time quantum walks
on the graph $G$ of the $n$-dimensional hypercube,
which was introduced and discussed by Moore and Russell~\cite{MooRus02RANDOM}.
Note that the $n$-dimensional hypercube can be
regarded as a Cayley graph of the Abelian group $\mathbb{Z}_2^n$ with
generators $g_1, g_2, \ldots, g_n$ such that
${g_1^2 = g_2^2 = \cdots = g_n^2 = e}$.
Thus, the shift
operator $S$ is defined as ${S\ket{a,v} = \ket{a,v \circ g_a}}$,
where $a \in \{1, \ldots, n\}$ is a label
and $v \in \mathbb{Z}_2^n$ is a vertex at which the particle is located.
Figure~\ref{Figure: hypercube} illustrates the case with the $3$-dimensional hypercube.

A coin-tossing operator $C$ is set as the Grover's diffusion operator~\cite{Gro96STOC}:
\[
 D =
 \left(
   \begin{array}{cccc}
     \tfrac{2}{n}-1 & \tfrac{2}{n}   & \dots  & \tfrac{2}{n}\\
     \tfrac{2}{n}   & \tfrac{2}{n}-1 & \dots  & \tfrac{2}{n}\\
     \vdots         & \vdots         & \ddots & \vdots\\
     \tfrac{2}{n}   & \tfrac{2}{n}   & \dots  & \tfrac{2}{n}-1
   \end{array}
 \right).
\]
Notice that coin-tossing operators in this model
should obey the permutation symmetry of
the hypercube.
Among such operators,
the Grover's diffusion operator is the one
farthest away from the identity operator under the standard operator norm
(see \cite{MooRus02RANDOM} for detailed discussions).
We call this quantum walk
a {\em symmetric walk on the $n$-dimensional hypercube\/}.
A number of properties were shown in~\cite{MooRus02RANDOM}
on the mixing time of this model in comparison with the classical case.

The process of this quantum walk is defined as follows:
\begin{enumerate}
 \item Initialize the system in the state $\ket{1,0}$.
       That is, let the particle be located at the vertex $0$ at time $0$,
       with the labeling being $1$.
 \item For every chosen number $t$ of steps, apply $W^t$ to the system,
       where ${W = S(D \otimes I_{\mathbb{Z}_2^n})}$,
       and then observe the location of the particle.
\end{enumerate}

\subsection{Numerical Simulations}

\begin{figure}[t]
\begin{center}
\includegraphics[height=5.75cm]{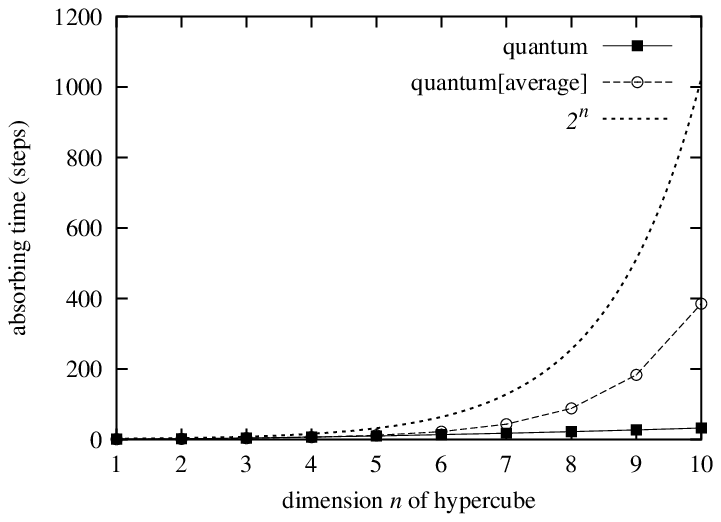}
\includegraphics[height=5.75cm]{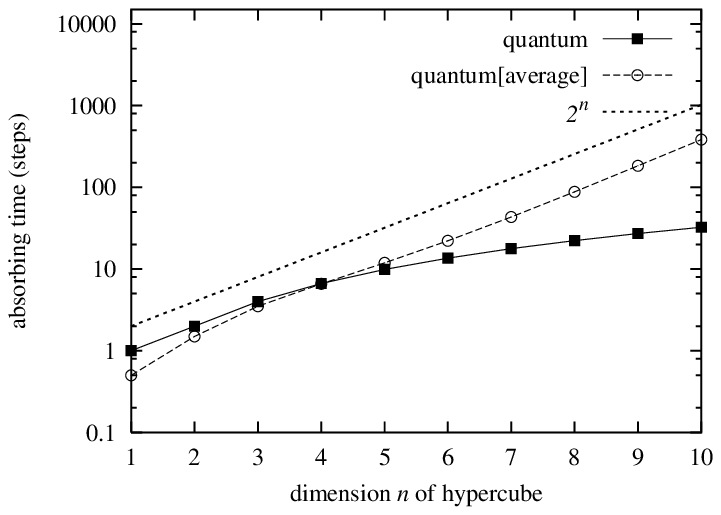}
\end{center}
\vspace*{-5mm}
\caption{Absorbing time of quantum symmetric walks on the $n$-dimensional hypercube.
The solid line ``quantum'' is the case that
absorbing vertex is located at $(1,1,\dots,1)$.
The dashed line ``quantum[average]'' is 
the case that absorbing vertex is located at random.
The dotted line represents the function $2^n$.}
\label{Figure: Absorbing time of quantum symmetric walks}
\end{figure}

Figure~\ref{Figure: Absorbing time of quantum symmetric walks} illustrates the relation
between the dimension $n$ of the hypercube (in $x$-axis)
and absorbing time of quantum walks (in $y$-axis).
One can see that,
if the absorbing vertex is located at random,
absorbing time averaged over all choices of the absorbing vertex
increases exponentially with respect to $n$.
This is similar to the classical case
for which the absorbing time is theoretically analyzed in the next section.
However,
if the absorbing vertex is located at $(1,1,\dots,1)$,
absorbing time seems to increase polynomially (quadratically, more precisely)
with respect to $n$.
This may suggest a striking contrast between quantum and classical
symmetric walks on the $n$-dimensional hypercube
that propagation between a particular
pair of vertices is {\em exponentially\/} faster in the quantum case.

Table~\ref{Table: Absorbing time of quantum symmetric walk} shows the relation between
(nominal and real) absorbing time of the quantum walk on the $8$-dimensional
hypercube and Hamming distance between the initial vertex and the
absorbing vertex.
One can see the following:
\begin{itemize}
\item
  absorbing probability is large if the absorbing vertex
  is located near the initial vertex
  or is located far from the initial vertex,
  while it is small if the absorbing vertex is located
  in the ``middle'' of the hypercube relative to the initial vertex,
\item
  nominal absorbing time has values of almost the same order
  except for the trivial case that Hamming distance is $0$,
\item
  nominal absorbing time is small, but real absorbing time is large,
  in the case of small absorbing probability.
\end{itemize}
In the classical case,
the shorter Hamming distance between the initial vertex and the
absorbing vertex is, the sooner the particle reaches the absorbing
vertex, which meets our intuition.
In the quantum case, however, our simulation result above is quite
counter-intuitive.

\begin{table}[t]
\begin{center}
\catcode`?=\active \def?{\phantom{0}}
\begin{tabular}{|c|ccccc|}
  \hline
  {\footnotesize Hamming} & {\footnotesize Absorbing} & & {\footnotesize Absorbing}
    & & {\footnotesize Absorbing}  \\
  {\footnotesize distance}& {\footnotesize time (real)} & & {\footnotesize time (nominal)}
    & & {\footnotesize probability}\\ \hline\hline
  0       & ?????0.0000???&=& ??0.0000?&/& ???1.0000???     \\ \hline
  1       & ????29.0000???&=& ?29.0000?&/& ???1.0000???     \\ \hline
  2       & ????59.0000???&=& ?16.8571?&/& ???0.2857???     \\ \hline
  3       & ????97.2444???&=& ?13.8921?&/& ???0.1429???     \\ \hline
  4       & ???115.5175???&=& ?13.2020?&/& ???0.1143???     \\ \hline
  5       & ????95.7844???&=& ?13.6835?&/& ???0.1429???     \\ \hline
  6       & ????56.3111???&=& ?16.0889?&/& ???0.2857???     \\ \hline
  7       & ????26.5603???&=& ?26.5603?&/& ???1.0000???     \\ \hline
  8       & ????22.3137???&=& ?22.3137?&/& ???1.0000???     \\
  \hline
\end{tabular}
\end{center}
\caption{Absorbing time and absorbing probability of a quantum symmetric walk
  on the $8$-dimensional hypercube.
}
\label{Table: Absorbing time of quantum symmetric walk}
\end{table}

From our simulation results illustrated by
Figure~\ref{Figure: Absorbing time of quantum symmetric walks} and Table~\ref{Table: Absorbing time of quantum symmetric walk},
we conjecture the following for the quantum case.

\begin{conjecture}
Absorbing probability of quantum walks on the $n$-dimensional hypercube is
$\min\Bigl\{1, \frac{n}{{n \choose i}}\Bigr\}$, where $i$ represents Hamming distance between the
initial vertex and the absorbing vertex.
\end{conjecture}
\begin{conjecture}
Nominal absorbing time of quantum walks on the $n$-dimensional hypercube is 
$O(n^2)$ independent of the location of the absorbing vertex,
except for the trivial case that the initial vertex is the absorbing vertex.
\end{conjecture}
\begin{conjecture}
Real absorbing time of quantum walks on the $n$-dimensional hypercube is
$\frac{n^2-n+2}{2}$ if Hamming distance between the
initial vertex and the absorbing vertex is $1$,
$O(n^2)$ if Hamming distance is $n$, and $\Theta(2^n)$ if Hamming distance is
close to $\frac{n}{2}$.
\end{conjecture}

\noindent
{\it Remark}~~
Kempe~\cite{Kem02quant-ph} recently showed that real absorbing time of
quantum walks on the $n$-dimensional hypercube is bounded polynomial in $n$
if Hamming distance between the initial vertex and the absorbing vertex is $n$.

\section{Theoretical Analyses of Symmetric Walks}
\label{Section: Theoretical Analyses}

\subsection{Classical Symmetric Walks}

First we state a property on the absorbing time
of classical symmetric walks on the $n$-dimensional hypercube.

\begin{proposition}
Absorbing time of classical symmetric walks on the $n$-dimensional
hypercube is $\Theta(2^n)$
independent of the location of the absorbing vertex,
except for the trivial case that the initial vertex is the absorbing vertex.
\label{Proposition: classical_2^n}
\end{proposition}

\begin{proof}
Recall that the particle doing a classical random walk determines a
direction to move randomly at each step.
Thus this walk obeys the
permutation symmetry of the hypercube and the vertices of the hypercube 
can be grouped in sets indexed by $i \in \{0, \ldots, n\}$,
each of which is a set 
of vertices whose Hamming distance from the initial vertex is $i$.

Let $s_i$ be absorbing time for the case that Hamming distance between
the initial vertex and the absorbing vertex is $i$.
Then, the following equations are satisfied:
\[
\begin{array}{ccl}
  s_0 & = &0,\\
  s_1 & = & \frac{1}{n}s_0 + \frac{n-1}{n}s_2 + 1,\\
  s_2 & = & \frac{2}{n}s_1 + \frac{n-2}{n}s_3 + 1,\\
  & \vdots &\\
  s_{n-1} & = & \frac{n-1}{n}s_{n-2} + \frac{1}{n}s_n + 1,\\
  s_n & = & s_{n-1} + 1,
\end{array}
\Longleftrightarrow
A \left(
    \begin{array}{c}
      s_0\\
      s_1\\
      s_2\\
      \vdots\\
      s_{n-1}\\
      s_n
    \end{array}
  \right)
=
\left(
  \begin{array}{c}
    0\\
    1\\
    1\\
    \vdots\\
    1\\
    1
  \end{array}
\right),
\]
where
\[
A = (a_{ij}) =
\left(
  \begin{array}{ccccccc}
    1 & 0 & \hspace*{8mm} & \hspace*{8mm} & \hspace*{8mm} & \hspace*{8mm} & \hspace*{8mm}\\
    -\frac{1}{n} & 1 & -\frac{n-1}{n}\\
    & -\frac{2}{n} & 1 & -\frac{n-2}{n}\\
    && \ddots & \ddots & \ddots\\
    &&& \ddots & \ddots & \ddots\\
    &&&& -\frac{n-1}{n} & 1 & -\frac{1}{n}\\
    \hspace*{8mm} & \hspace*{8mm} &&&& -1 & 1
  \end{array}
\right).
\]

Let a matrix $B = (b_{ij})$ such that
\[
b_{ij} = \begin{cases}
  1 & \mbox{if $j = 1$},\\
  0 & \mbox{if $i = 1, j \ge 2$},\\
  {n \choose {j-1}} \sum_{l=0}^{\min(i,j)-2}\frac{1}{{{n-1} \choose l}} & \mbox{if $i,j \ge 2$},
\end{cases}
\]
and let $c_{ik} = \sum_{j=1}^{n+1} a_{ij}b_{jk}$.
After some calculations, we have the following.
\begin{itemize}
\item In the case $i = 1$,
  $c_{ik} = b_{1k} = 1$ if $k = 1$, otherwise $c_{ik} = 0$.
\item In the case $i = n+1$,
  $c_{ik} = -b_{nk}+b_{n+1,k} = 1$ if $k = n+1$, otherwise $c_{ik} = 0$.
\item In the case $2 \le i \le n$,
  $c_{ik} = -\frac{i-1}{n}b_{i-1,k} + b_{ik} - \frac{n-i+1}{n}b_{i+1,k} = 1$ if $k = i$, otherwise $c_{ik} = 0$.
\end{itemize}
Therefore, $AB$ is the identity matrix, and thus $B = A^{-1}$.
It follows that
\[
s_{i-1} = \sum_{j=2}^{n+1}b_{ij}
=
\begin{cases}
  0 & \mbox{if $i = 1$},\\
  \sum_{j=2}^{n+1}
  \left({n \choose {j-1}}\sum_{l=0}^{\min(i,j)-2}
    \frac{1}{{{n-1} \choose l}}
  \right) & \mbox{if $i \geq 2$}.
\end{cases}
\]
From this, it is obvious that $s_i$ increases monotonously with
respect to $i$.
For $i \ge 1$, we have
\[
  2^n - 1 = \sum_{j=2}^{n+1}{n \choose {j-1}} \le s_i
  < 3\sum_{j=2}^{n+1}{n \choose {j-1}} = 3(2^n - 1),
\]
since $1 \le \sum_{l=0}^{\min(i,j)-2}\frac{1}{{{n-1} \choose l}} < 3$.
Thus we have the assertion.
\end{proof}

This result might be counterintuitive in some sense,
since the vertices at small Hamming distance, say 1, from the initial vertex
may seem easy to find.
However, the following will give an intuitive explanation of this behavior:
after one step,
the probability of going away from the absorbing vertex is $1-1/n$,
which is much higher than the probability $1/n$ of finding it.

\subsection{Quantum Symmetric Walks}

In our simulation results in the previous section,
quantum symmetric walks on the hypercube
behave in a manner quite different from classical ones.
In particular,
it is remarkable that in the quantum case propagation
between a particular pair of vertices
seems exponentially faster than in the classical case.
Here we try to describe absorbing time of
quantum symmetric walks on the $n$-dimensional hypercube
as a function of $n$.
The results in this subsection
do not give a simple characterization of
the absorbing time of the quantum walk on the hypercube,
but at least simplify its numerical calculation considerably.

The following lemma states that
we do not need to keep track of the amplitudes of all basis
states for symmetric quantum walks on the hypercube.
\begin{lemma}
\label{Lemma: Un}
For a symmetric quantum walk on the $n$-dimensional hypercube
with the initial vertex $o$,
define sets
$\{F_i\}$ and $\{B_i\}$ as follows:
\begin{eqnarray*}
F_i &=& \bigl\{(a,v) \mid \Hamming{v,o}=i,\;\Hamming{v \circ g_a,o}=i+1 \bigr\},\\
B_i &=& \bigl\{(a,v) \mid \Hamming{v,o}=i,\;\Hamming{v \circ g_a,o}=i-1 \bigr\},
\end{eqnarray*}
where $\Hamming{x,y}$ denotes the Hamming distance between $x$ and $y$.
Let two unit vectors $\ket{f_i}$ and $\ket{b_i}$
in $\mathcal{H}_A \otimes \mathcal{H}_V$ be defined by
\[
\begin{array}{lcll}
\ket{f_i} &=& \displaystyle\frac{1}{\sqrt{|F_i|}} \sum_{(a,v) \in F_i} \ket{a,v}
            & ~~\mbox{\rm for $i = 0, \ldots, n-1$},\\
\ket{b_i} &=& \displaystyle\frac{1}{\sqrt{|B_i|}} \sum_{(a,v) \in B_i} \ket{a,v}
            & ~~\mbox{\rm for $i = 1, \ldots, n$}.
\end{array}
\]
 Then, a transition step of symmetric quantum walks is given
by the following $2n \times 2n$ unitary matrix:
\[
U_n =
\left(
  \begin{array}{cccccccccc}
    0 & -\frac{n-2}{n} & \frac{\sqrt{4n-4}}{n} \\
    1 & 0 & 0 \\
    0 & 0 & 0 & -\frac{n-4}{n} & \frac{\sqrt{8n-16}}{n} \\
    & \frac{\sqrt{4n-4}}{n} & \frac{n-2}{n} & 0 & 0 \\
    &&& 0 & 0 & -\frac{n-6}{n} & \ddots \\
    &&& \frac{\sqrt{8n-16}}{n} & \frac{n-4}{n} & \ddots & \ddots \\
    &&&&& \ddots & \ddots & \frac{n-2}{n} & \frac{\sqrt{4n-4}}{n} \\
    &&&&& \ddots & -\frac{n-4}{n} & 0 & 0 & 0 \\
    &&&&&&& 0 & 0 & 1 \\
    &&&&&&& \frac{\sqrt{4n-4}}{n} & -\frac{n-2}{n} & 0
  \end{array}
\right)
\]
with the order of bases
$\ket{f_0}, \ket{b_1}, \ket{f_1}, \ldots, \ket{b_{n-1}}, \ket{f_{n-1}}, \ket{b_{n}}$.
\end{lemma}

\begin{proof}
First, we calculate $|F_i|$ and $|B_i|$.
The number of vertices $v$ satisfying $\Hamming{v,o} = i$ is ${n \choose i}$,
and for such $v$, the number of labels $a$ satisfying ${\Hamming{v \circ g_a,o} = i+1}$ is $n-i$.
Therefore, the number of basis vectors in $F_i$ is $(n-i){n \choose i}$.
Similarly, the number of basis vectors in $B_i$ is $i{n \choose i}$.

Now, after applying a coin-tossing operator
(which is the Grover's diffusion operator) to $\ket{f_i}$,
we have
\begin{eqnarray*}
(C \otimes I) \ket{f_i}
&=&
\frac{1}{\sqrt{|F_i|}} \sum_{(a,v) \in F_i}
  \Biggl(
    \frac{2}{n}\sum_{1 \le b \le n}\ket{b,v} - \ket{a,v}
  \Biggr)\\
&=&
\frac{1}{\sqrt{|F_i|}}
  \Biggl(
    \frac{2(n-i)}{n}\sum_{(a,v) \in B_i}\ket{a,v}
      + \frac{n - 2i}{n}\sum_{(a,v) \in F_i}\ket{a,v}
  \Biggr).
\end{eqnarray*}
Then, the shift operator is applied to this vector to have
\begin{eqnarray}
S(C \otimes I) \ket{f_i}
&=&
\frac{1}{\sqrt{|F_i|}}
  \Biggl(
    \frac{2(n-i)}{n}\sum_{(a,v) \in F_{i-1}}\ket{a,v}
      + \frac{n - 2i}{n}\sum_{(a,v) \in B_{i+1}}\ket{a,v}
  \Biggr)
\nonumber\\
&=&
\sqrt{\frac{|F_{i-1}|}{|F_i|}} \cdot \frac{2(n-i)}{n} \ket{f_{i-1}}
  + \sqrt{\frac{|B_{i+1}|}{|F_i|}} \cdot \frac{n - 2i}{n} \ket{b_{i+1}}
\nonumber\\
&=&
\frac{\sqrt{4ni-4i^2}}{n}\ket{f_{i-1}} + \frac{n-2i}{n}\ket{b_{i+1}}.
\label{Equation: SC|fi>}
\end{eqnarray}

Similarly, 
the coin-tossing operator and shift operator are applied in sequence
to the state $\ket{b_i}$ to have 
\begin{equation}
S(C \otimes I) \ket{b_i}
=
-\frac{n-2i}{n}\ket{f_{i-1}} + \frac{\sqrt{4ni-4i^2}}{n}\ket{b_{i+1}}.
\label{Equation: SC|bi>}
\end{equation}
The lemma holds immediately from (\ref{Equation: SC|fi>}) and (\ref{Equation: SC|bi>}).
\end{proof}

The following is immediate from Lemma~\ref{Lemma: Un}.

\begin{corollary}
One step of transition of symmetric quantum walks on the $n$-dimensional
hypercube with the absorbing vertex located at $(1, 1, \ldots, 1)$ is
described by a CP map
$\Lambda: \rho_t \mapsto \rho_{t+1} =
  U_nP_n^\prime\rho_tP_n^\prime U_n^{\dagger} + I_nP_n\rho_tP_nI_n$,
where
$\rho_t$ and $\rho_{t+1}$ are density operators of the system at time
$t$ and $t+1$, respectively, $I_n$ is the $2n \times 2n$ identity
matrix, $P_n$ is a $2n \times 2n$ projection matrix 
whose $(2n,2n)$-element is 1 and all the others are 0,
and $P_n^\prime = I_n - P_n$.
\end{corollary}

\begin{figure}[t]
\begin{center}
\includegraphics[height=5.75cm]{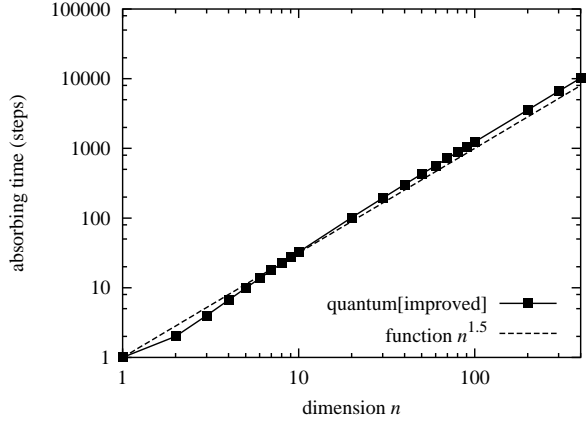}
\end{center}
\vspace*{-5mm}
\caption{Absorbing time of quantum walks on higher dimensional
hypercubes with the absorbing vertex located at $(1, 1, \ldots, 1)$.
The dotted line represents the function $n^{1.5}$.}
\label{Figure: larger n}
\end{figure}

By this corollary, we can do numerical simulations for much larger $n$,
say $n=500$.
Figure~\ref{Figure: larger n} illustrates the relation
between the dimension $n$ of the hypercube (in $x$-axis)
and absorbing time of quantum walks (in $y$-axis) for larger $n$.
One can see that absorbing time is close to $1.25 n^{1.5}$.

In what follows, we focus on the case that the absorbing vertex
is at $(1, 1, \ldots, 1)$.

Since both absorbing probability and nominal absorbing time are defined
as power series, and time evolution of the quantum walk is described by
a CP map (hence by a linear operator),
it is significant to investigate the properties of the matrix corresponding
to time evolution operator and its eigenvalues
in order to study behaviors of absorbing time with respect to the dimension $n$.
First, we prove the following lemma.
\begin{lemma}
Every eigenvalue of $U_nP_n^\prime$ has its absolute value of less then $1$,
where $P_n^\prime = I_n - P_n$.
\label{Lemma: eigenvalue}
\end{lemma}

\begin{proof}
It is obvious that
every eigenvalue of $U_nP_n^\prime$ has
its absolute value of at most $1$.

Suppose that there exists an eigenvalue $\lambda$ of
$U_nP_n^\prime$ such that $|\lambda| = 1$.
Let $\boldsymbol{e}$ be
the eigenvector of $U_nP_n^\prime$ corresponding to $\lambda$.
Then the $2n$th element of $\boldsymbol{e}$ is $0$ and
$P_n^\prime \boldsymbol{e} = \boldsymbol{e}$,
since applying the
projection operator $P_n^\prime$ changes the $2n$th element of
$\boldsymbol{e}$ into $0$
and $|\lambda| = 1$.
Thus we have $U_n \boldsymbol{e} = \lambda\boldsymbol{e}$,
which implies that $\boldsymbol{e}$ is an eigenvector of $U_n$.
However, it is easy to check by calculation that,
in every eigenvector of $U_n$,
the $2n$th element of it is not $0$, which is a contradiction.
\end{proof}

By using Lemma~\ref{Lemma: eigenvalue},
exact values of
absorbing probability and absorbing time
can be described for quantum walks on
the $n$-dimensional hypercube with the absorbing vertex located at
$(1, 1, \ldots, 1)$.

\begin{proposition}
Let $X_n$ be the $2n \times 2n$ symmetric matrix satisfying
${X_n - U_nP_n^\prime X_nP_n^\prime U_n^{\dagger} = \rho_0}$,
where
$P_n^\prime = I_n - P_n$
and
$\rho_0$ is the initial density operator of
a $2n \times 2n$ matrix whose $(1,1)$-element is $1$
and all the other elements are $0$.
Then the absorbing time of quantum walks
on the $n$-dimensional hypercube with the
absorbing vertex located at $(1, 1, \ldots, 1)$ is $\tr{X_n} - 1$.
\label{Proposition: X_n}
\end{proposition}

\begin{proof}
Let $p(t)$ denote the probability that the particle reaches
the absorbing vertex $(1, 1, \ldots, 1)$ at time $t$ for the first time.
Then we have
\begin{eqnarray*}
p(t)
& = &
\tr \bigl(
      (U_n P_n^\prime)^t\rho_0(P_n^\prime U_n^{\dagger})^t
      P_n
    \bigr)
\\
& = &
\tr \bigl(
      (U_n P_n^\prime)^t\rho_0(P_n^\prime U_n^{\dagger})^t
        - (U_n P_n^\prime)^{t+1}\rho_0(P_n^\prime U_n^{\dagger})^{t+1}
    \bigr).
\end{eqnarray*}
From Lemma~\ref{Lemma: eigenvalue},
both absorbing probability
and nominal absorbing time are convergent series.
Thus we have the following.
\[
\begin{array}{l}
  \begin{array}{l}
    \prob = \displaystyle\sum_{t=0}^\infty p(t)
          = \displaystyle\sum_{t=0}^\infty
              \tr{
                \bigl(
                  (U_n P_n^\prime)^t\rho_0(P_n^\prime U_n^{\dagger})^t
                    - (U_n P_n^\prime)^{t+1}\rho_0(P_n^\prime U_n^{\dagger})^{t+1}
                \bigr)
              }
    \\
    \hspace*{0.83cm}
      = \tr{\rho_0} = 1,
  \end{array}
  \\
  \\
  \begin{array}{l}
    \nominal = \displaystyle\sum_{t=0}^\infty t p(t)
             = \displaystyle\sum_{t=0}^\infty t
                 \tr{
                   \bigl(
                     (U_n P_n^\prime)^t\rho_0(P_n^\prime U_n^{\dagger})^t
                       - (U_n P_n^\prime)^{t+1}\rho_0(P_n^\prime U_n^{\dagger})^{t+1}
                   \bigr)
                 }
    \\
    \hspace*{0.83cm}
      = \tr{
          \Biggl(
            \displaystyle\sum_{t=0}^\infty
              (U_n P_n^\prime)^{t+1} \rho_0 (P_n^\prime U_n^{\dagger})^{t+1}
          \Biggr)
        }
      = \tr(U_n P_n^\prime X_n P_n^\prime U_n^{\dagger}) = \tr{X_n} - 1.
 \end{array}
\end{array}
\]
It follows that real absorbing time is $\tr{X_n} - 1$.
\end{proof}

Another characterization of values of the absorbing time
is given by the following proposition.

\begin{proposition}
Let $f_n(x)$ be a power series of $x$ defined as
\[
f_n(x) = \sum_{t=0}^\infty a_t x^t
       = \frac{2^{n-1} (n-1)!}{n^{n-1}}
         \cdot
         \frac{x^{n+1}}{\det(xI_n - U_nP_n^\prime)}.
\]
Then the absorbing time of quantum walks on the $n$-dimensional hypercube
with the absorbing vertex located at $(1, 1, \ldots, 1)$ is
$\sum_{t=0}^\infty t|a_t|^2$.
\end{proposition}

\begin{proof}
Let $\bra{b_n} = (0~\cdots~0~1)$
and $\ket{f_0} = (1~0~\cdots~0)^T$.
Then $\bra{b_n}(U_nP_n^\prime)^t\ket{f_0}$ gives the amplitude
that the particle reaches
the absorbing vertex $(1, 1, \ldots, 1)$ at time $t$ for the first time.
Consider a power series of $x$,
${g_n(x) = \sum_{t=0}^\infty \bra{b_n}(U_nP_n^\prime)^t\ket{f_0} x^t}$.
We prove that this $g_n(x)$ is equivalent to $f_n(x)$.

Note that
${g_n(x)
   = (0~\cdots~0~1) (\sum_{t=0}^\infty (x U_nP_n^\prime)^t)(1~0~\cdots~0)^T}$,
which is equal to the $(2n,1)$-element of $(I_n-xU_nP_n)^{-1}$.
Thus we have for $x \neq 0$,
\[
g_n(x)
  = \frac{x^{2n}}{\det\bigl(xI_n-U_nP_n^\prime\bigr)}
       (-1)^{2n+1}\varDelta_{1,2n},
\]
where $\varDelta_{1,2n}$ is a minor of $x I_n-U_nP_n^\prime$ with
respect to $(1,2n)$.

It is straightforward to show that
$\varDelta_{1,2n} = -\frac{2^{n-1}(n-1)!}{n^{n-1}} \cdot \frac{1}{x^{n-1}}$,
and thus
\[
g_n(x) = \frac{2^{n-1}(n-1)!}{n^{n-1}}
         \cdot
         \frac{x^{n+1}}{\det\bigl(xI_n-U_nP_n^\prime\bigr)} = f_n(x).
\]

By the definition of $g_n(x)$, each coefficient $a_t$ of $x^t$
in the power series $f_n(x)$ corresponds to the amplitude that
the particle reaches the absorbing vertex $(1, 1, \ldots, 1)$ at time $t$ for the first time.
Since the absorbing probability is $1$ from the proof
of Proposition~\ref{Proposition: X_n},
the real absorbing time is
$\real = \sum_{t=0}^\infty t|a_t|^2$.
\end{proof}

\section{Conclusions}
\label{Section: Conclusions}

This paper focused on the absorbing probability and time of quantum walks
through numerical simulations and theoretical analyses
on Hadamard walks on the line
and symmetric walks on the hypercube.

In our numerical simulations,
quantum walks behaved in manners quite different from classical walks. 
In particular, quantum walks on the hypercube appeared exponentially faster
than classical ones in absorbing time under certain situations.
For some of our conjectures based on these numerical results,
recent papers~\cite{Kem02quant-ph, BacCopGolJoyWat02quant-ph}
subsequent to ours have given theoretical proofs.

As for our theoretical analyses on symmetric walks on the hypercube,
the authors believe that they are quite useful in simplifying
numerical calculations of the absorbing time of them.

\section*{Acknowledgements}

The authors would like to thank Jun Hasegawa and Takashi Yamada
for their help in our numerical simulations.
The authors would also like to thank anonymous referees for their
helpful comments.

\end{document}